\documentstyle{article}

\newcommand{\CPN}{{\rm CP}^{N-1}}

\newcommand\simle{\rlap{\raise 2pt \hbox{$<$}}{\lower 2pt \hbox{$\sim$}}}

\title{\rightline{\tenrm IFUP-TH 44/92}\bigskip
Topological susceptibility and string tension in ${\CPN}$ models
\footnote{Talk presented at the {\em Lattice '92\/} Conference, Amsterdam.}
}

\author{%
Massimo Campostrini, Paolo Rossi, and Ettore Vicari\\
\tenit Istituto Nazionale di Fisica Nucleare, Sezione di Pisa,\\
\tenit Dipartimento di Fisica dell'Universit\`a, I-56126 Pisa, Italy.%
}

\date{October 6th, 1992}

\begin{document}

\maketitle

\begin{abstract}
We determined the topological susceptibility and the string tension
in the lattice ${\rm CP}^{N-1}$ models for a wide range of values
of $N$, in particular for $N=4,10,21,41$.
Quantitative agreement with the large-$N$ predictions is found
for the ${\rm CP}^{20}$ and the ${\rm CP}^{40}$ models.
\end{abstract}

\section{INTRODUCTION}

The most attractive feature of two dimensional ${\rm CP}^{N-1}$ models
is their similarity with the Yang-Mills theories in four
dimensions. Most properties of ${\rm CP}^{N-1}$ models have been obtained
in the context of the $1/N$ expansion around the large-$N$ saddle point
solution.
An alternative and more general non-perturbative approach
is the simulation of the theory on the lattice. It is then important
to perform Monte Carlo simulations of the lattice ${\rm CP}^{N-1}$
models to check the validity
and the range of applicability of the $1/N$ expansion.

The large-$N$ expansion predicts an exponential area law
behavior for sufficiently large Wilson loops, which implies confinement,
with a string tension $O(1/N)$,
and absence of screening due to the dynamical
matter fields.
${\rm CP}^{N-1}$ models have a non-trivial topological structure;
at large $N$ the topological susceptibility turns out
to be $O(1/N)$.
In this talk the features of ${\rm CP}^{N-1}$ models concerning
confinement and topology are investigated.
In order to study the approach to the large-$N$ asymptotic regime,
we performed numerical simulations for a wide range of values of $N$,
in particular $N=4,10,21,41$.
A more detailed discussion of these simulations and of their results
can be found in Refs.\ \cite{CPNlatt_2,CPNlatt_3,CPN_MP}.

\section{LATTICE FORMULATION}

We regularize the theory on the lattice by considering
the following action:
\begin{equation}
S_{\rm g} = -N\beta\sum_{n,\mu}\left(
   \bar z_{n+\mu}z_n\lambda_{n,\mu} +
   \bar z_nz_{n+\mu}\bar\lambda_{n,\mu} \right),
\label{basic}
\end{equation}
where $z_n$ is an $N$-component complex scalar field, satisfying
$\bar z_nz_n = 1$,
and $\lambda_{n,\mu}$ is a ${\rm U}(1)$ gauge field.
We also considered its tree Symanzik improved counterpart
$S_{\rm g}^{\rm Sym}$ in order to test universality.

The standard correlation length $\xi_{\rm w}$ is extracted from the
long-distance behavior of the zero space momentum correlation
function $G_P$
of two projector operators $P_{ij}(x) = \bar z_i(x) z_j(x)$.
We define an alternative correlation length $\xi_G$
from the second moment of the correlation function $G_P$.
For $N=2$ $\xi_G/\xi_{\rm w}\simeq 1$ within 1\% \cite{CPNlatt},
while the large-$N$ expansion predicts \cite{CPN}
\begin{equation}
{\xi_{\rm G}\over\xi_{\rm w}}=\sqrt{{2\over 3}}\,+\,
O\left( {1\over N^{2/3} } \right)\;.
\label{ratioxi}
\end{equation}
Unlike $\xi_{\rm w}$, which is
a non-analytic function of $1/N$ around $N=\infty$, $\xi_{\rm G}$
can be expand in powers of $1/N$ \cite{CPN}.

\section{TOPOLOGY}

The large-$N$ predictions concerning the topological susceptibility
$\chi_t$ are \cite{CPN-lett}
\begin{equation}
\chi_{\rm t}\xi_G^2 = {1\over 2\pi N}\,\left(1-{0.3801\over N}\right) +
O\left({1\over N^3}\right)\,,
\label{chipred}
\end{equation}
and \cite{Luscher-lett}
\begin{equation}
\chi_{\rm t}\xi_{\rm w}^2 = {3\over 4\pi N}\,+\,
O\left({1\over N^{5/3}}\right)\,.
\label{chipredluscher}
\end{equation}
Eq.\ (\ref{chipred}) and Eq.\ (\ref{chipredluscher}) are not in
contradiction with each other due to Eq.\ (\ref{ratioxi}),
but the first one should be testable at lower values of $N$ according
to the powers of $N$ in the neglected terms.

A troublesome point in the lattice simulation technique is the study
of the topological properties and the determination of $\chi_t$.
While for large $N$
the geometrical definition $\chi_t^g$
\cite{Berg-Luscher} is expected to reproduce
the physical topological susceptibility,
at low $N$ $\chi_t^g$ could receive unphysical contributions
from exceptional configurations, called dislocations.
The dislocation contributions may either survive in the continuum limit,
as it happens for the ${\rm CP}^1$
or $O(3)$ $\sigma$ model, or push the scaling region for $\chi_t^g$
to very large $\beta$ values.

Another approach, the field theoretical method, relies
on a definition of topological charge density by a local polynomial
in the lattice variables $q_L$.
The correlation at zero momentum of two $q^L$ operators $\chi_t^L$
is related to $\chi_t$ through the following equation:
\begin{eqnarray}
\chi_{\rm t}^L(\beta) &=& a^2 Z(\beta)^2\chi_{\rm t} \,+\, a^2 \,A(\beta)
   \langle S(x)\rangle \nonumber \\
&+&  P(\beta)\langle I \rangle \,+ \,O(a^4) \,.
\label{chilchi}
\end{eqnarray}
$S(x)$ is the trace of the energy-momentum tensor, $I$ is the
identity operator.
$Z(\beta)$, $P(\beta)$, and $A(\beta)$ are
ultraviolet effects, since they originate from the ultraviolet
cutoff-dependent modes.
The field theoretical method consists in measuring
$\chi_{\rm t}^L(\beta)$, evaluating
$Z(\beta)$, $A(\beta)$ and $P(\beta)$, and using Eq.\ (\ref{chilchi})
to extract $\chi_{\rm t}$.

A third method, called cooling method, measures $\chi_t$
on an ensemble of configurations cooled by locally minimizing the
action \cite{Teper}.

\subsection{The ${\rm CP}^3$ model}

We determined $\chi_t$ by performing simulations with both
actions $S_{\rm g}$ and $S_{\rm g}^{\rm Sym}$ and by using different
methods.
In Fig.~1 the dimensionless quantity $\chi_t \xi_G^2$ is
plotted versus $\xi_G$.
Data obtained by the geometrical method,
indicated by the filled symbols in Fig.~1,
violate universality, showing
that $N=4$ is not large enough to suppress the unphysical
configurations contributing to $\chi_t^g$, at least for $\xi \le 30$.

To apply the field theoretical method,
we considered the following lattice operator:
\begin{equation}
q^L(x) = -{i\over 2\pi}\sum_{\mu\nu} \epsilon_{\mu\nu}
   {\rm Tr}\left[ P(x)\Delta_\mu P(x)
   \Delta_\nu P(x) \right]\,,
\label{localq1}
\end{equation}
where $\Delta_\mu$ is a symmetrized version of the finite
derivative.
We neglected the contribution of the mixing with $S(x)$;
this assumption is supported by perturbative arguments.
We obtained non-perturbative estimates of $Z(\beta$ and $P(\beta)$
by using the ``heating method'' described in Ref.\ \cite{chi-letter}.
This method relies on the distinction between the fluctuations
at $l\sim a$, contributing to the renormalizations, and those
at $l\sim \xi$ determining the relevant topological properties.
Due to the critical slow down phenomenon,  fluctuations
at $l\sim a$ are soon thermalized, while, for large $\xi$, the topological
charge thermalization is much slower, allowing a direct determination of
$Z(\beta)$ (when heating an instanton configuration)
and $P(\beta)$ (when heating the flat configuration).

An independent measure of $\chi_t$ is obtained using the cooling method.
The results of both field theoretical and cooling
methods are shown in Fig.~1. Scaling,
universality, and good agreement between the two methods are observed.

In conclusion, we quote for the ${\rm CP}^3$ model $\chi_t\xi_G^2\simeq 0.06$
with an uncertainty of about 10\%.

\subsection{The ${\rm CP}^9$ model}

We measured $\chi_t$ by using the geometrical definition.
Data were taken for $\xi_G$ up to about 10 lattice units.
Data for $S_{\rm g}$ show a slow
approach to scaling, while for $S_{\rm g}^{\rm Sym}$ a
better behavior is observed.
For $S_{\rm g}$ the leading scaling
violation term must be $O(\ln \xi/\xi^2)$ when $\xi \to \infty$
\cite{Symanzik}.  For the tree Symanzik improved actions the
leading logarithmic corrections are absent, and scale violations are
$O(\xi^{-2})$ \cite{Paffuti}.
In Fig.~2 we plot
$\chi_{\rm t}^{\rm g}\xi_G^2$ versus $\ln \xi_{\rm G}/\xi_{\rm G}^2$.
Assuming that the scaling violation
term proportional to $\ln\xi/\xi^2$ is already dominant in our range
of correlation lengths, we extrapolated data of $\chi_{\rm t}^{\rm g}$
for the action $S_{\rm g}$.  We found
$\chi_{\rm t}^{\rm g}\xi_G^2 = 0.0174(12)$, which is
in agreement with the
value of $\chi_{\rm t}^{\rm g}\xi_G^2$ obtained with the action
$S_{\rm g}^{\rm Sym}$.
We then conclude that for the ${\rm CP}^9$
model $\chi_{\rm t}^{\rm g}$ is a good estimator of the topological
susceptibility.

\subsection{The ${\rm CP}^{20}$ and ${\rm CP}^{40}$ models}

As shown in Fig.~3
for both the ${\rm CP}^{20}$ and ${\rm CP}^{40}$ models,
data of $\chi_t$, obtained by using the geometrical method,
are consistent with the large-$N$ prediction
(\ref{chipred}),
whose results are indicated by the dashed lines.
Fitting the data for the ${\rm CP}^{20}$ model, we found
$\chi_t \xi_G^2\,=\,0.0076(3)$,
to be compared with the value $\chi_t \xi_G^2\,=\,0.00744$
coming from Eq.\ (\ref{chipred}).
However,
the above result still disagrees with
Eq.\ (\ref{chipredluscher}), which would require
$\xi_G^2/\xi_{\rm w}^2\simeq 2/3$, while we found
$\xi_G^2/\xi_{\rm w}^2\simeq 0.91$ in our simulations.

In Fig.~4 we summarize our results
by plotting $\chi_t\xi_G^2$ versus $1/N$, showing how the
topological susceptibility approaches the large-$N$ asymptotic behavior
(\ref{chipred}), represented by the solid line.

\section{CONFINEMENT}

The large-$N$ prediction for the string tension $\sigma$ is
\begin{equation}
\sigma\xi_G^2\,=\,{\pi\over N} \,+\,O\left( {1\over N^2}\right)\;.
\label{sigmapred}
\end{equation}

The string tension can be easily extracted by measuring the Creutz
ratios defined by
\begin{equation}
\chi(R,T) = \ln {W(R,T{-}1)\,W(R{-}1,T)
   \over W(R,T)\,W(R{-}1,T{-}1)} \,.
\end{equation}
where $W(R,T)$ are the Wilson loops constructed with the $\lambda_\mu$
field.
In a 2-d finite lattice with periodic boundary conditions, the
large abelian Wilson loops of a confining theory are subject to large
finite size effects.
For sufficiently large $R$ the behavior
of the Creutz ratios $\eta(R)\equiv \chi(R,R)$,
i.e. of those with equal arguments,
should be \cite{CPNlatt_2}:
\begin{equation}
\eta(R) \simeq \sigma\,\left[ 1-\left({2R-1\over L}\right)^2\right]\;,
\label{Crratcpn}
\end{equation}
where $L$ is the lattice size.
To compare data from different lattices it is convenient
to define a rescaled Creutz ratio
\begin{equation}
\eta_{\rm r}(R) = \eta(R)\,
    \left[ 1-\left({2R-1\over L}\right)^2\right]^{-1}
\simeq \sigma\;.
\label{Crratres}
\end{equation}

In Fig.~5 we plot $\eta_{\rm r}(R)\xi_G^2$ versus
the physical distance $r=R/\xi_G$ for the ${\rm CP}^9$,
${\rm CP}^{20}$, and ${\rm CP}^{40}$ models.
Starting from $r\simeq 2$ the rescaled Creutz ratios show
a clear plateau which is the evidence of the string tension.
A similar behavior is observed for $N=4$.
We do not see evidence of screening effects (at least up to $r\simeq
3\xi_G$, data for larger $r$ were too noisy) confirming the picture coming
from the $1/N$ expansion.
Data for the ${\rm CP}^{20}$ and ${\rm CP}^{40}$ show a good agreement
with the large-$N$ prediction (\ref{sigmapred}), represented by the dashed
lines in Fig.~5.

\section{CONCLUSION}

With regard to topology and confinement, numerical results
show a qualitative agreement with the continuum $1/N$
expansion for all values of $N$, while
quantitative agreement is found
for $N=21$ and for higher values of $N$.
On the other hand, the approach to the large-$N$
asymptotic regime of quantities involving the mass gap appears
very slow and the ${\rm CP}^{40}$ model should be still outside
the region where the complete mass spectrum predicted
by the $1/N$ expansion could be observed \cite{CPNlatt_3}.
The agreement in this sector of the theory is expected to be reached
at very large $N$, because of the large coefficient in the effective
expansion parameter $6\pi/N$ that can be extracted from a non-relativistic
Schr\"odinger equation analysis of the linear confining
potential \cite{CPN}.

\newpage
\def\thesection{}
\section{Figure captions}
\bigskip\par\noindent
{\bf Fig.~1: }$\chi_t\xi_G^2$ versus $\xi_G$ for the ${\rm CP}^3$ model.
We plot data of the geometrical method (for $S_{\rm g}$ ``$\Box$'' and
$S_{\rm g}^{\rm Sym}$ ``$\Diamond$''), the field theoretical method
($S_{\rm g}$ ``$\Diamond$'', $S_{\rm g}^{\rm Sym}$ ``$\times$''),
and  the cooling method
($S_{\rm g}$ ``$\Box$'', $S_{\rm g}^{\rm Sym}$
``$\scriptstyle\bigcirc$'').
\bigskip\par\noindent
{\bf Fig.~2: }$\chi_t\xi_G^2$ versus $\ln\xi_G/\xi_G^2$ for the
${\rm CP}^9$ model.
\bigskip\par\noindent
{\bf Fig.~3: }$\chi_t\xi_G^2$ for the
${\rm CP}^{20}$ and the ${\rm CP}^{40}$ models.
\bigskip\par\noindent
{\bf Fig.~4: }$\chi_t\xi_G^2$ versus $1/N$.
\bigskip\par\noindent
{\bf Fig.~5: }$\eta_r(R)\xi_G^2$ versus $r=R/\xi_G$,
for $N=10$, $N=21$, and $N=41$. For each $N$ we show data at two values
of $\beta$.
\end{document}